\documentclass[fleqn,twoside]{article}
\usepackage{espcrc2,epsf}

\def\openone{\leavevmode\hbox{\small1\normalsize\kern-.33em1}}

\def\slash{\!\!\!\!/}

\def\a0limit{a \rightarrow 0}

\def\openone{\leavevmode\hbox{\small1\normalsize\kern-.33em1}}

\def\spose#1{\hbox to 0pt{#1\hss}}
\def\ltapprox{\mathrel{\spose{\lower 3pt\hbox{$\mathchar"218$}}
 \raise 2.0pt\hbox{$\mathchar"13C$}}}
\def\gtapprox{\mathrel{\spose{\lower 3pt\hbox{$\mathchar"218$}}
 \raise 2.0pt\hbox{$\mathchar"13E$}}}
\def\inapprox{\mathrel{\spose{\lower 3pt\hbox{$\mathchar"218$}}
 \raise 2.0pt\hbox{$\mathchar"232$}}}

\def\figsizea{3.0}
\def\figsizeb{2.0}

\def\figure#1#2#3#4{\epsfxsize=#4 truein
\medskip
\centerline{\epsffile{#1}}
\bigskip
\centerline{\vbox{{\bf \noindent Figure #2.} #3}}
\smallskip 
}

\def\one{Two flavor dynamical Wilson HMC $\Phi$  $\beta = 5.2$, $\kappa$=0.18, 
V= $32^3 \times 64$.
}
\def\two{An abstraction of the BG/L node chip.
}
\def\three{The 2,048 node, 11.5 Teraflops BlueGene/L prototype at the IBM T.J. Watson lab.
}
\def\four{Performance vs. local lattice size.
}
\def\five{Performance of the Wilson CG inverter vs. $\#$  of CPUs
for a $4^3 \times 16$ local lattice. 
}

\newcommand{\AmS}{{\protect\the\textfont2
  A\kern-.1667em\lower.5ex\hbox{M}\kern-.125emS}}

\title{QCD on the BlueGene/L Supercomputer}

\author{
    G. Bhanot,
    D. Chen,
    A. Gara,
    J. Sexton,
    and P. Vranas \thanks{Speaker.}
    \address{
    IBM T.J. Watson Research Center, 
    Route 134,
    Yorktown Heights, NY 10598, USA.
    }
}

\begin{document}

\begin{abstract}
In June 2004 QCD was simulated for the first time at sustained speed 
exceeding 1 TeraFlops in the BlueGene/L supercomputer at the IBM T.J. 
Watson Research Lab. The implementation and performance of QCD in 
the BlueGene/L is presented.
\end{abstract}

\maketitle  

\section{Introduction}
\label{sec:introduction}
Lattice gauge theory has been intimately connected to numerical
simulations and computer hardware from the
beginning. More specifically lattice field theories lend themselves
naturally to numerical simulations on massively parallel
supercomputers. These machines have defined the landmarks for
computational speed through the years. There have only been a few and
can all be remembered by name and by the new window they opened to
research in field theory.

\figure{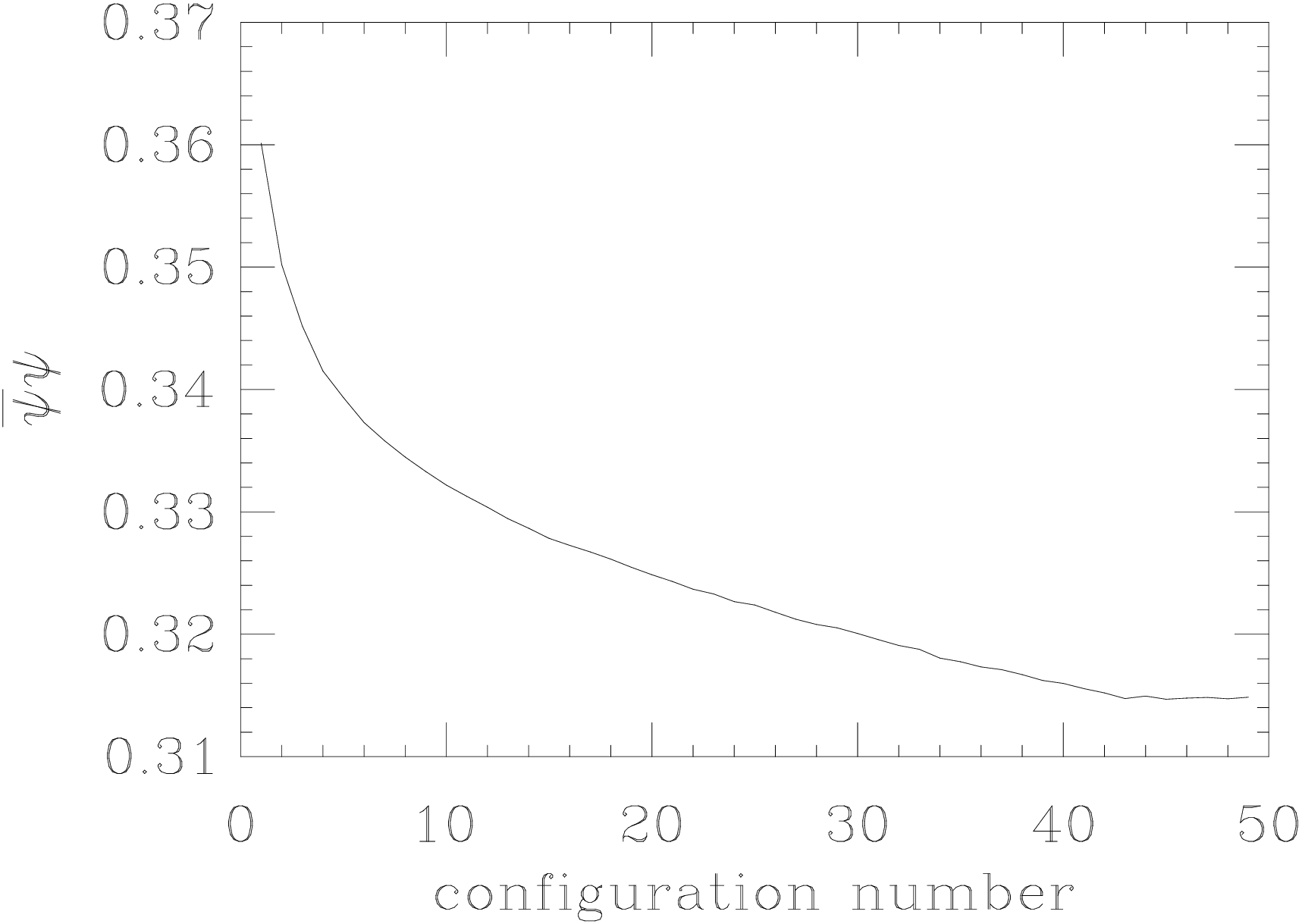}{1}{\one}{\figsizeb}

In this sense, when in June 2004 lattice QCD ran on a 1024 node 5.6
Teraflop prototype BlueGene/L (BG/L) supercomputer at the IBM
T.J. Watson Research lab for eight hours, at sustained speed exceeding
1 Teraflops, another landmark was crossed signaling the next generation
of lattice gauge theory calculations. That simulation was dynamical
QCD with Wilson fermions on a $32^3 \times 64$ lattice at $\beta=5.2$
and $\kappa=0.180$. Figure 1 shows the thermalization of the chiral
condensate from an ordered gauge configuration. Furthermore, using the
full 2048 node 11.5 Teraflop prototype another dynamical run was
done. It sustained more than 2 Teraflops for 3 hours ($V=64^3 \times
16$, $\beta=5.1$ and $\kappa=0.180$). At the moment of this writing,
the available QCD code can sustain up to about $19\%$ of peak.

For details about BG/L see for example \cite{Chen}.  For other QCD
related supercomputers see for example
\cite{other_supers}. The largest planned BG/L installations will be a
115 Teraflops 20K node machine at IBM Watson and a 367 Teraflops 64K
node machine at Lawrence Livermore National Lab in 2005.
In the June 2004 ranking of the world's fastest 
supercomputers a BG/L machine (located at IBM Rochester) rated number 4 
and another one (located at IBM Watson) rated number 8.

Programming Lattice QCD to run efficiently on these large machines has
always been a challenge. Here we describe the relevant BG/L hardware
and how QCD is coded for it. We also present performance and scaling
measurements.

\section{The microchip}
\label{sec:2}
Figure 2 shows an abstraction of the BG/L chip (node).  The chip has
two embedded IBM 440 CPU cores. Each core has an enhanced floating
point unit capable of two multiply/add (MADD) instructions per cycle.
Therefore the chip is capable of a peak of 8 floating point
instructions per cycle.  At the operation speed of 700 MHz the chip
delivers 5.6 GFlops of peak speed.  The memory hierarch is shown in
figure 2.  

\figure{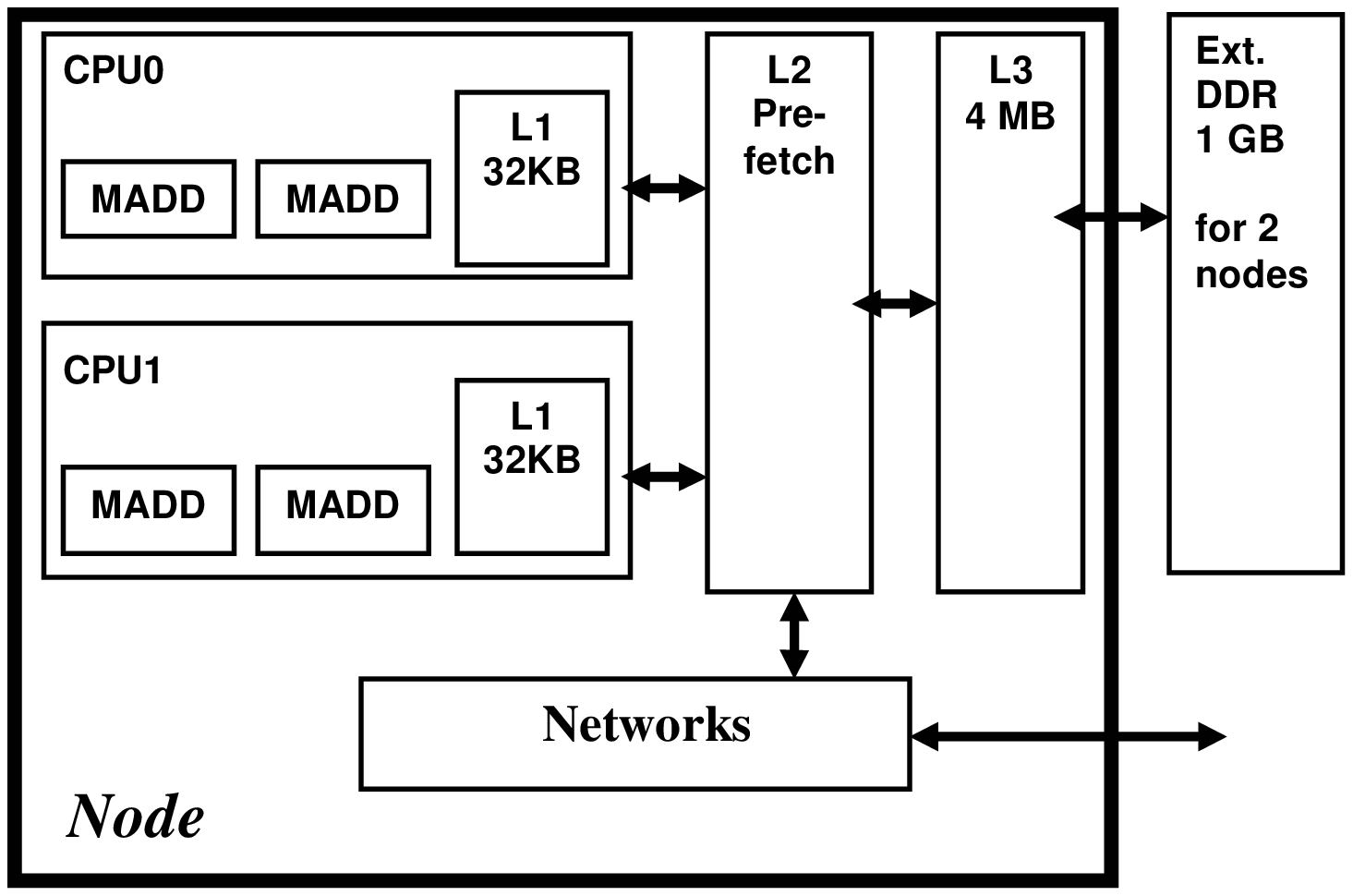}{2}{\two}{\figsizea}

Several network controllers are built in on the chip.  The
main network connects the nodes in a 3-dimensional torus grid with
nearest neighbor high speed interconnects.  The hardware implements a
sophisticated dynamic virtual cut-through network. Packets can be sent
from any node to any other node without CPU intervention. Furthermore,
the nodes are also connected via a global tree network that can
efficiently perform global operations as well as file I/O.  Also, the
nodes are connected via separate interrupt and control networks.
The system is fully symmetric with respect to the two CPU cores.

\section{The BlueGene/L supercomputer}
\label{sec:1}
Two chips are assembled on a compute-card. Sixteen compute-cards are
connected with connectors on a node-card. In case of a chip failure
the compute-card can be easily replaced. Thirty-two node-cards are
plugged into 2 vertical midplane cards. The 1024 node assembly is
housed in a ``refrigerator'' sized cabinet (rack). Many racks can be
connected with cables to form large systems. Figure 3 shows the 2
rack prototype at the IBM Watson lab. Also, a single rack can be
configured as an $8\times 8 \times 16$ torus or as two $8 \times 8
\times 8$ torus grids without having to be re-cabled.

\section{QCD on the hardware}
\label{sec:3}
QCD can use the two CPU cores of the BG/L chip in two basic modes.  1)
Co-processor mode where CPU0 does all the computations and CPU1 does
all the communications (including MPI etc.). The 4th direction is
internal to CPU0.  Communications can overlap with computations but
the peak performance is then limited to $5.6/2 = 2.8$
GFlops. 2). Virtual node mode where CPU0 and CPU1 act as independent
``virtual nodes'', each with its own memory space.  Each one does both
computations and communications. The 4th direction is along the two
CPUs that  communicate via common memory. Computations
and communications can not overlap but the peak performance is
the full 5.6 GFlops.

\figure{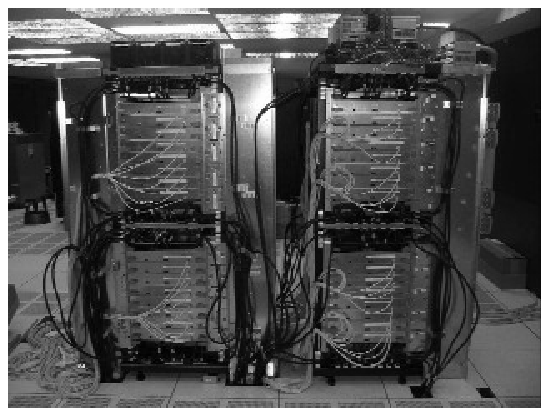}{3}{\three}{\figsizea}

\section{The Wilson operator}
\label{sec:4}
The Wilson operator is coded in virtual node mode. The standard spin
projection and reconstruction algorithm is used.  The multiplication of
the gauge field with a spinor is done using the spin projected two
component spinors. Also, only two component spinors need to be
transferred between nodes.  All computations use the double
multiply/add instructions. Computations overlap with
load/stores. Local performance is bounded by memory access to L3.
Communications do not overlap with computations or memory access.
Because of the above there is an interesting tradeoff: For small local
size one has fast L1 memory access but more communications (larger
surface to volume ratio). For large local size one has slower L3
memory access but less communications. The inner most kernel is
written in ``pseudo-assembly'' (c or c++ inline assembly).

\figure{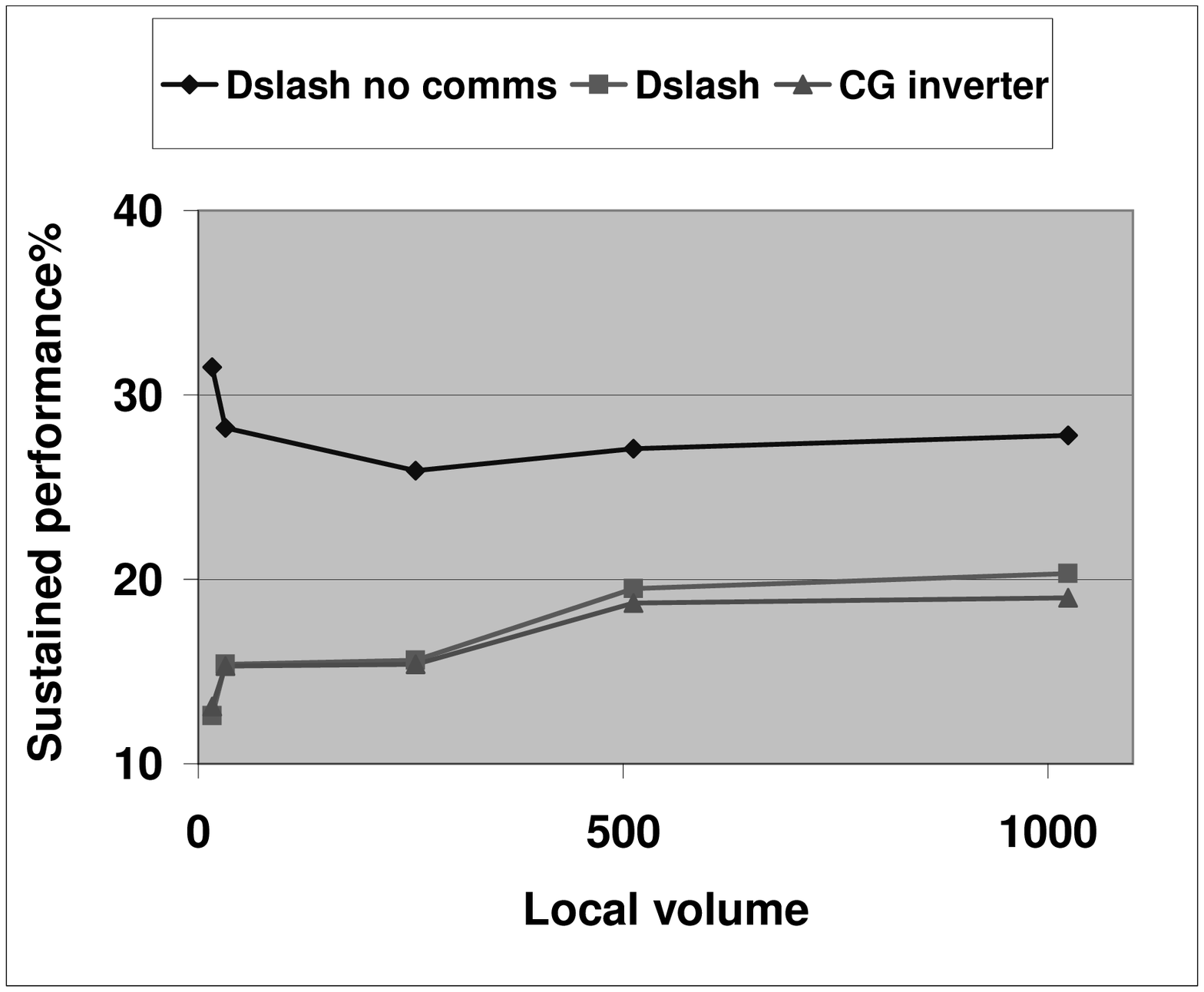}{4}{\four}{\figsizea}

\section{Performance}
\label{sec:5}
Figure 4 shows the performance of $D \slash$ without communications,
of $D \slash$ with communications and of the full inverter. Figure 5
shows the performance for a fixed $4^3 \times 16$ local lattice as a
function of increasing machine size (here the CG
inverter global sums were done on the torus).

\section{Dream machines}
\label{sec:6}
Here are some examples of interesting machine configurations. Of
course this is just a sample of what is possible. A full 1024 node
rack, $(8, 8, 16, 2)$ CPUs, with a local lattice of $(4, 4, 4, 16)$
sites gives a lattice of $(32,32,64,32)$ sites. This is appealing for
next generation dynamical zero temperature calculations.  A half node
rack, 512 nodes, $(8, 8, 8, 2)$ CPUs with a local lattice of $(4, 4,
4, 4)$ sites gives a lattice of $(32,32,32,8)$ sites. This is suitable
for next generation thermodynamic calculations.

If 16 racks of the BG/L system at IBM Watson are configured as 
$(16, 32, 32, 2)$ and a local lattice with $(4, 2, 2, 32)$ sites is used
then one has a lattice with $(64,64,64,64)$ sites. If the 64 racks of
the BG/L system at LLNL is configured as $(64, 32, 32, 2)$ and a local
lattice of $(2, 2, 2, 32)$ is used then one has a lattice with
$(128,64,64,64)$ sites. Clearly these are dream machines for QCD.

\figure{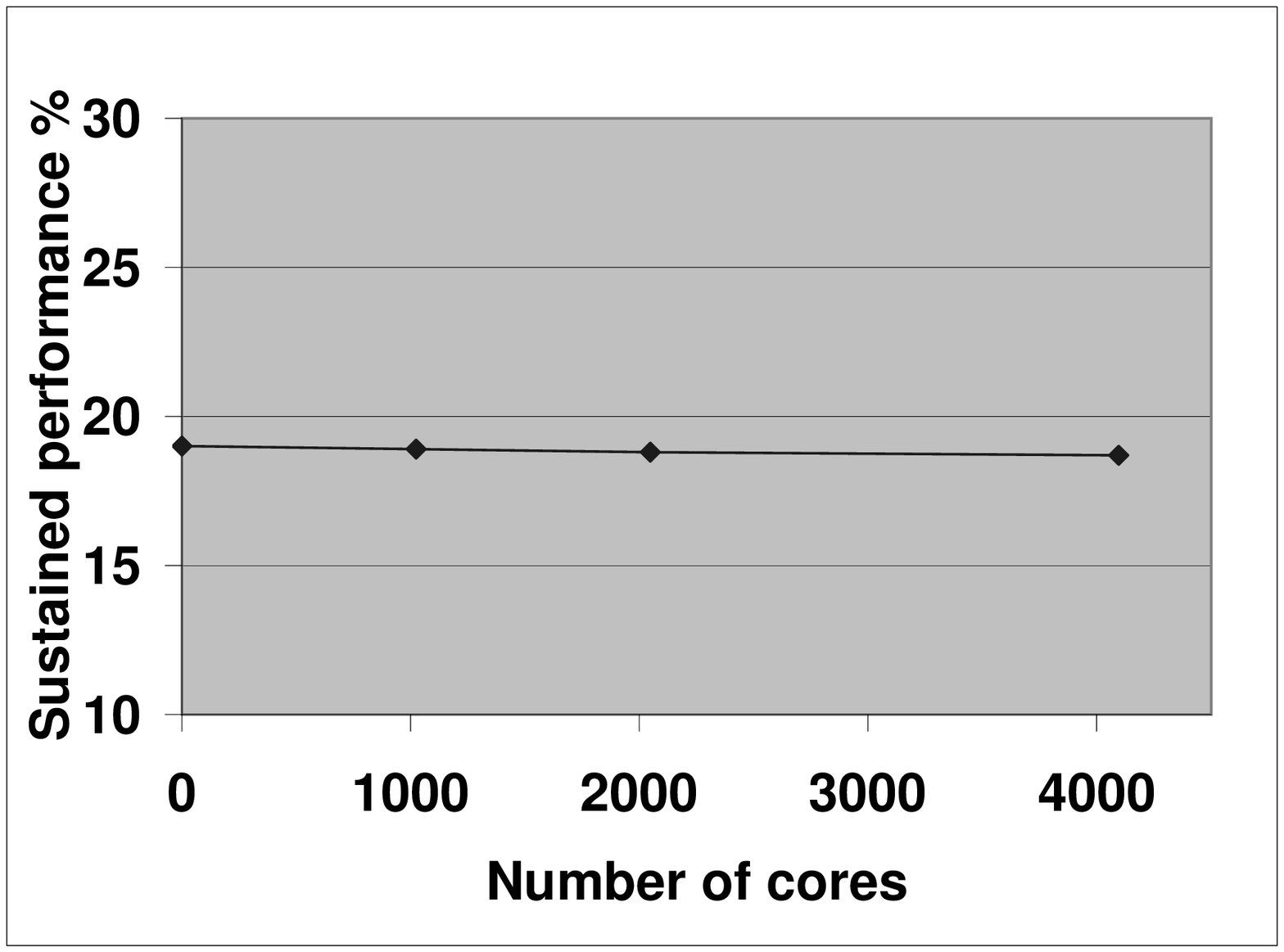}{5}{\five}{\figsizea}

\section{Physics software}
\label{sec:7}
Most of the high level physics code is the Columbia C++ physics system
(cps). The full system ported easily and worked immediately.

\section{Conclusions}
\label{sec:8}
QCD crossed the 1 sustained-Teraflops landmark in June 2004.
In the next year, because of analytical and supercomputer developments,
dynamical QCD will likely get to L/a = 32 at physical quark masses and
perhaps even more...

\section*{Acknowledgments}
We would like to thank the QCDOC collaboration for useful discussions
and for providing us with the Columbia physics system software.


\end{document}